\documentstyle[sprocl,epsfig]{article}

\def\be{\begin{equation}}
\def\ee{\end{equation}}
\def\bea{\begin{eqnarray}}
\def\eea{\end{eqnarray}}
\begin{document}

\hfill hep-ph/0410121

\hfill IPPP-04-52

\hfill DCPT-04-104

\title{SUSY SEARCHES IN COMBINED LHC/LC ANALYSES\\}

\author{G. MOORTGAT-PICK$^{1}$\footnote{Speaker at LCWS04},
K. DESCH$^{2}$, J. KALINOWSKI$^{1,3}$,
\\ M.M. NOJIRI$^{4}$, 
G. POLESELLO$^{5}$}

\address{$^{1}$ IPPP, University of Durham, DH1 3LE Durham, UK\\
$^{2}$ Institut f\"ur Experimentalphysik, Hamburg University, 22607 Hamburg, Germany\\
$^{3}$ Institute of Theoretical Physics, Warsaw University, 00681 Warsaw, Poland\\
$^{4}$ YITP, Kyoto University, Kyoto 606 8502, Japan\\
$^{5}$ INFN, Sezione di Pavia,  Pavia 27100, Italy
}


\maketitle\abstracts{
We present a case study for the
synergy of combined LHC and LC analyses in Susy searches in which
simultaneous running of both machines is very important.
In this study only
light non-coloured Susy particles are accessible at a Linear
Collider with an initial energy of $\sqrt{s}=500$~GeV. Nevertheless
the precise analysis at the LC 
leads to an
accurate Susy parameter determination and prediction of
heavy Susy particles. Providing these LC results as input for the LHC analyses
could be crucial for the identification of signals
resulting in a direct measurement of the heavy neutralinos.
The interplay of the LHC and LC will thus 
provide an important consistency test of the
underlying model.
}

%***********************************************************************
\section{Introduction}
%***********************************************************************
In the
Minimal Supersymmetric Standard Model (MSSM) one is faced with
around 105 new free parameters. They have to be precisely determined 
at future experiments in order to
reveal the underlying structure of the model.
Due to the clear signatures at the Linear Collider (LC)
a largely model-independent determination of masses,
couplings, mixing angles, phases and quantum numbers of kinematically
accessible particles can be done. 
Assuming the LHC running to overlap with the first stage of a 
LC with $\sqrt{s}=500$~GeV (LC$_{500}$),
significant impact of LC$_{500}$ on the 
LHC analyses is expected. For the reference point SPS1a\cite{SPS} we
show how measurements of  the light chargino/neutralino states at the LC$_{500}$ 
may help identifying signals 
of the heavy
neutralinos produced at the LHC and consequently 
improve their mass measurement.
Conversely, the LHC results will increase the precision of Susy
parameter determination at the LC. 
The interplay of both colliders will therefore 
provide a powerful
consistency check of the model 
and may outline future strategies for new physics searches at the LHC
\cite{DKMNP}. 

%%%%%%%%%%%%%%%%%%%%%%%%%%%%%%%%%%%%%%%%%%%%%
\section{Susy Analyses At The LHC And The LC}
\begin{figure}[ht]
\begin{center}
     \vspace*{.2cm}
\epsfig{figure=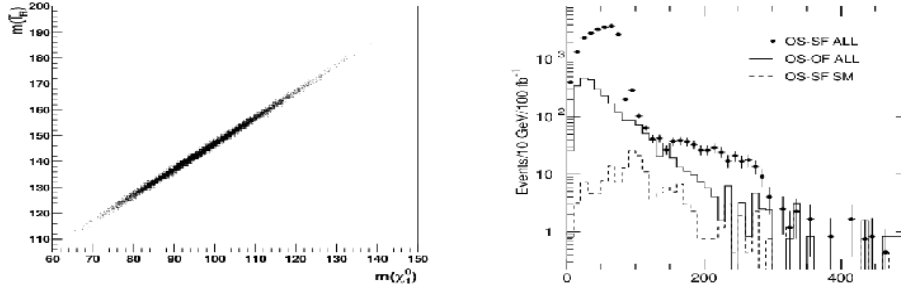,height=1.5in,width=4.7in}
\end{center}
\caption{Mass measurements at the LHC$^{3,4}$: correlation of the
  fitted $m_{\tilde{\ell}_R}$ and 
LSP $m_{\tilde{\chi}^0_1}$ in the scenario SPS1a (left), and
invariant mass spectrum of the heavy neutralino/chargino decay chains$^5$ (right).
The dilepton OS-SF lepton edge of $\tilde{\chi}^0_4$ is the edge between
200~GeV$<m_{ll}<$400~GeV.
\label{fig-lhc}}
\end{figure}

%%%%%%%%%%%%%%%%%%%%%%%%%%%%%%%%%%%%%%%%%%%%%
{\bf LHC Susy studies:}\\
Detailed simulations of the LHC capabilities for the reference point SPS1a\cite{SPS}
were  carried out~\cite{LHCstudy}; the masses of the Susy particles
can be studied by analysing complicated decay chains, like
$\tilde{q}_L\to \tilde{\chi}^0_2 q \to \tilde{\ell}^{\mp}_R \ell^{\pm}
q\to \tilde{\chi}^0_1 
\ell^{\mp} \ell^{\pm} q$ and $\tilde{q}_L\to \tilde{\chi}^0_4 q \to
\tilde{\ell}^{\mp}_R \ell^{\pm} 
q\to \tilde{\chi}^0_1 
\ell^{\mp} \ell^{\pm} q$, which might be difficult to resolve, in
particular the latter one. 
A joint fit of various kinematic 'edges'
yields an overconstraint  
system and leads to an indirect knowledge on the mass of the
lightest Susy particle (LSP)
with, however, some assumptions  
about particle identities. 
We
show in  Fig.~\ref{fig-lhc} (left)  the strong correlation between the
fitted values of $m_{\tilde{\ell}_R}$
and  $m_{\tilde{\chi}^0_1}$. 
The precise reconstruction of the 
states in the decay chains requires in particular the knowledge of $m_{\tilde{\chi}^0_1}$.  
In the SPS1a reference point  the heavy charged and the neutral
gaugino/higgsino particles 
can lead to identical final states and the association
of the edges 
to the corresponding sparticles
is difficult.
In combination with measured invariant masses one can derive the
Opposite-Sign Same-Flavour 
(OS-SF) signal of the heavy particle with $\delta(m)=5.1$~GeV, and
under specific assumptions one 
can interpret the edge in Fig.~\ref{fig-lhc} (right) as that of the
$\tilde{\chi}^0_4$ particle \cite{LHCstudy,Giacomo}.\\
%%%%%%%%%%%%%%%%%%%%%%%%%%%%%%%%%
{\bf LC$_{500}$ Susy studies:}\\
%%%%%%%%%%%%%%%%%%%%%%%%%%%%%%%%%
The precise measurement of the  masses
as well as the production cross sections
of the light non-coloured sparticles alone may lead to a rather
precise determination of the fundamental 
Susy parameters which govern the chargino-neutralino sector 
(see e.g. \cite{ckmz} and references therein).
The masses of the heavier neutralinos and the heavier chargino can then be
predicted.

In order to determine the parameters we exploit only $m_{\tilde{\chi}^{\pm}_1}$,
$m_{\tilde{\chi}^0_{1,2}}$ and the corresponding cross section measurements at a LC
with both beams polarised and take into account
the simulated errors of all relevant masses\cite{martyn}, 
$m_{\tilde{\chi}^{\pm}_1}$, $m_{\tilde{\chi}^0_{1,2}}$
 $m_{\tilde{e}_{L,R}}$, $m_{\tilde{\nu}}$ , 1$\sigma$ statistical
errors for the polarised 
cross sections (on a basis of 100~fb$^{-1}$ per each polarisation configuration), a 
polarisation uncertainty of $\Delta P(e^{\pm})/P(e^ {\pm})=0.5\%$ and estimated  
systematic errors \cite{DKMNP}.

Within the allowed error bars we derive
a very accurate determination of the underlying
fundamental Susy parameters, see Table~\ref{tab_par_com} (2nd line)
and get the following predictions for the heavier particles:
\begin{equation}
m_{\tilde{\chi}^0_3}=359.2\pm8.6~\mbox{GeV},\quad
m_{\tilde{\chi}^0_4}=378.2\pm 8.1~\mbox{GeV},\quad 
m_{\tilde{\chi}^{\pm}_2}=378.8\pm 7.8
\label{eq_pred}
\end{equation}
%%%%%%%%%%%%%%
{\bf Combined LHC+LC$_{500}$ studies:}\\
%%%%%%%%%%%%%%
Feeding the results of the LC analysis as input into the LHC
analysis improves the LHC analysis in several ways:\\
$\bullet$ increase of statistical sensitivity due to the mass
predictions ('look elsewhere effect'), which
could be crucial for the search for statistically marginal
signals;\\ 
$\bullet$ clear identification of the dilepton edge from
the $\tilde{\chi}^0_4$ decay chain, followed by an accurate measurement
of $m_{\tilde{\chi}^0_4}=377.87\pm 2.23$~GeV;\\ 
$\bullet$
better accuracy also for $m_{\tilde{\chi}^0_2}$
(e.g. $\delta(m_{\tilde{\chi}^0_2})=0.08$~GeV)  due
the precise knowledge of the LSP mass, $m_{\tilde{\chi}^0_1}$.\\
The precise
identification of a dilepton edge right at the predicted mass with the
help of the LC means an important check of the
underlying Susy model.

Using these improved results from the LHC analysis as input for further
analyses at the LC leads also to an improvement in the Susy parameter
determination, see Table~\ref{tab_par_com}, 3rd line.

%%%%%%%%%%%%%%%%%%%%%%
\section{Conclusions}
%%%%%%%%%%%%%%%%%%%%%%
The analysis has been performed within the general framework
of the unconstrained MSSM. 
We focused on the situation where only the light states
($\tilde{\chi}^0_1$, $\tilde{\chi}^0_2$, $\tilde{\chi}^\pm_1$) are
accessible at the first stage of the LC, as exemplified in the SPS1a scenario.
The masses of heavier chargino and neutralinos
can subsequently be predicted which in turn 
lead to an increase of statistical sensitivity and to a clear
identification of the heavy 
particles in the corresponding decay chains at the LHC analysis.
Feeding back the LHC results into further analysis at the LC$_{500}$
leads to an even more accurate model-independent 
determination of the Susy parameters.
Such a combined study provides therefore a sensitive test of the model.

%%%%%%%%%%%%%%%%%%%
\begin{table}[ht]
     \begin{center}
\begin{tabular}{|c|cccc|} \hline
& $M_1$ & $M_2$ & $\mu$ & $\tan\beta$  \\
\hline
\multicolumn{1}{|c|}{input} & 99.1 & 192.7 & 352.4 & 10 \\
LC$_{500}$& $99.1\pm 0.2$& $192.7\pm 0.6$& $352.8\pm 8.9$ &
$10.3\pm 1.5$ \\
LHC+LC$_{500}$ & $99.1\pm 0.1$ & $192.7\pm 0.3$ &
{$352.4\pm 2.1$} & {$10.2\pm 0.6$} \\ \hline
\end{tabular}
\end{center}
\caption{Susy parameters with 1 $\sigma$ errors derived from
the LC data collected at the first phase of operation, and from 
the combined analysis of the
LHC and LC$_{500}$ data.
\label{tab_par_com}}
\end{table}

\section*{Acknowledgements} The work is supported in part by the
  European Commission 5-th Framework Contract HPRN-CT-2000-00149. 
JK has partially been supported by the KBN
  grant 2 P03B 040 24 (2003-2005).

%%%%%%%%%%%%%%%%%%%%%%%%%%%%


\section*{References}
\begin{thebibliography}{99}
\bibitem{DKMNP}{K.~Desch, J.~Kalinowski, G.~Moortgat-Pick,
  M.~M.~Nojiri and G.~Polesello, 
%``SUSY parameter determination in combined analyses at LHC/LC,''
JHEP {\bf 0402} (2004) 035 [hep-ph/0312069];
%%CITATION = HEP-PH 0312069;%%
see also in hep-ph/0402295;
G.~Moortgat-Pick,
%``Synergy of combined LHC and LC analyses in SUSY searches,''
hep-ph/0406180.}
%%CITATION = HEP-PH 0402295;%%
\bibitem{SPS}{B.~C.~Allanach {\it et al.},
%``The Snowmass points and slopes: Benchmarks for SUSY searches,''
%in {\it Proc. of the APS/DPF/DPB Summer Study on the
%Future of Particle Physics (Snowmass 2001) } ed. N.~Graf,
Eur.\ Phys.\ J.\ C {\bf 25} (2002) 113
[eConf {\bf C010630} (2001) P125]
[hep-ph/0202233];
  N.~Ghodbane and \mbox{H.-U.~Martyn},
%``Compilation of SUSY particle spectra from Snowmass 2001 benchmark
%models,''
  hep-ph/0201233;
%%CITATION = HEP-PH 0201233;%%
see also  http://www.cpt.dur.ac.uk/$\tilde{}$ georg/sps/.}
%%%%%
\bibitem{LHCstudy}{The ATLAS Collaboration, ATLAS TDR 15, CERN/LHCC/99-15 (1999);
B.~C.~Allanach, C.~G.~Lester, M.~A.~Parker and B.~R.~Webber,
%``Measuring sparticle masses in non-universal string inspired models at  the
%LHC,''
JHEP {\bf 0009} (2000) 004
[hep-ph/0007009];
B.K. Gjelsten, J. Hisano, K. Kawagoe,
E. Lytken, D. Miller, M. Nojiri, P. Osland, G. Polesello;
M. Chiorboli, A. DeRoeck, B.K. Gjelsten, K. Kawagoe, E. Lytken, H.-U. Martyn,
 D. Miller, P. Osland, G. Polesello and A. Tricomi,
contributions in the LHC/LC working group document, see also
http://www.ippp.dur.ac.uk/$\tilde{}$ georg/lhclc/.}
%%%%%
\bibitem{Giacomo}{G. Polesello, 
%{\it Prospects for the detection of heavy charginos and 
%neutralinos with the ATLAS detector at the LHC}, 
J. Phys. G: Nucl. Part. Phys. {\bf 30} 
No 9 (2004) 1185-1199.}
%%%%%
\bibitem{martyn}
H.U. Martyn, LC-note LC-PHSM-2003-071 and hep-ph/0406123; M. Dima et al., Phys. Rev.
{\bf D65} (2002) 071701; M. Ball, diploma thesis, University of Hamburg, 2003. 
%%%%%
\bibitem{ckmz}{S.~Y.~Choi, J.~Kalinowski, G.~Moortgat-Pick and P.~M.~Zerwas,
%``Analysis of the neutralino system in supersymmetric theories,''
Eur.\ Phys.\ J.\ C {\bf 22} (2001) 563
[Addendum-ibid.\ C {\bf 23} (2002) 769]
[hep-ph/0108117, hep-ph/0202039].}
%%CITATION = HEP-PH 0108117;%%
%%CITATION = HEP-PH 0202039;%%


\end{thebibliography}
\end{document}